# Two Circular-Rotational Eigenmodes in Vortex Gyrotropic Motions in Soft Magnetic Nanodots

Ki-Suk Lee and Sang-Koog Kim*

Research Center for Spin Dynamics and Spin-Wave Devices, Seoul National University, Seoul 151-744, Republic of Korea

Nanospintronics Laboratory, Department of Materials Science and Engineering, College of Engineering, Seoul National University, Seoul 151-744, Republic of Korea

We found, by micromagnetic numerical and analytical calculations, that the clockwise (CW) and counterclockwise (CCW) circular-rotational motions of a magnetic vortex core in a soft magnetic circular nanodot are the elementary eigenmodes existing in the gyrotropic motion with respect to the corresponding CW and CCW circular-rotational-field eigenbasis. Any steady-state vortex gyrotropic motions driven by a linearly polarized oscillating in-plane magnetic field in the linear regime can be perfectly understood according to the superposition of the two circular eigenmodes, which show asymmetric resonance characteristics reflecting the vortex polarization. The relative magnitudes in the amplitude and phase between the CCW and CW eigenmodes determine the elongation and orientation of the orbital trajectories of the vortex core motions, respectively, which trajectories vary with the polarization and chirality of the given vortex as well as the field frequency across the resonance frequency.

Periodic oscillatory phenomena are common in nature, existing in diverse physical systems such as oscillating electrons or atoms driven by electromagnetic waves, vibrating molecules, a mechanically vibrating mass coupled to a spring and simple pendulum, and, in the case of nonmechanical oscillation, electric circuits [1]. These oscillators have the common characteristic of free or damped oscillation, with the corresponding natural frequencies. Under oscillating driving forces of the oscillators' characteristic natural frequencies, such oscillators are resonantly excited with extremely large amplitudes of motion, which phenomenon is called the resonance effect [1]. A most promising example in magnetic systems is the dynamic behavior of a magnetic vortex (MV), interest in which has been increasing rapidly owing to its nontrivial static and dynamic properties [2,3]. The MV structure consists of in-plane curling magnetizations (**M**s) and out-of-plane **M**s at its core region, the so-called vortex core (VC), which is known to be a ground state in geometrically confined soft magnetic elements of micron size or smaller [2]. When, in such a confined system, magnetic fields (or currents) with harmonic oscillations or pulses are applied to the vortex, the VC rotates around its equilbrium position at a characteristic eigenfrequency of several hundred MHz [4-9]. The responsible force is the gryroforce exerting on the VC, which is in balance with the restoring force due mainly to the long-range dipole-dopole interaction dominating in a confined magnetic element [5]. Such vortex excitation is known to be the translation mode or gyrotropic motion in the dot plane. It is known that the rotation sense of such gyrotropic motion is determined by the polarization $p$ of the vortex, represented by the **M** orientation of the VC [$p=$ +1(-1) for up(down)-core orientation]. If the frequency of an oscillating field or current is close to the vortex eigenfrequency, the VC motion is resonantly excited [9,10]. More recently, the resonant VC motion has attracted much attention on account of its related ultrafast VC switching applicable to information storage [11-16]. In addition, the variation of the circular and elliptical shapes of the orbital trajectories of the on- and off-resonance VC motions, driven by a linearly polarized oscillating magnetic field (LPH), was reported to vary with the field frequency across the vortex eigenfrequency, but has yet been clearly understood, since the true eigenmodes of the vortex gyrotropic motions remain incompletely understood. In this Letter, having considered the results of the present theoretical and numerical studies, we posit that these VC motions can be clearly understood by considering them to be the superposition of counterclockwise (CCW) and clockwise (CW) circular-rotational eigenmodes, and also by considering their aysmmetric resonance effect. Additionally, we report that the relative magnitudes in the amplitude and phase between the two circular eigenmodes determine complex vortex motions, providing information on how they vary with the polarization and chilarity of the given vortex state as well as the field frequency.

In the present study, we employed micromagnetic numerical simulations of vortex **M** dynamics using the OOMMF code [18] that utilizes the Landau-Lifshitz-Gilbert (LLG) equation of motion $\partial \mathbf{M}/\partial t = -\gamma \left(\mathbf{M} \times \mathbf{H}_{eff}\right) + \alpha/M_s \left(\mathbf{M} \times \partial \mathbf{M}/\partial t\right)$ [19] with the phenomenological damping constant $\alpha$, the gyromagnetic ratio $\gamma$, $M_s = |\mathbf{M}|$, and the effective field $\mathbf{H}_{eff}$. Also, we carried out analytical calculations of the linear-regime VC motions [8], based on the linearized Thiele's equation of motion (see Ref. [20] for details). As a model system, we chose a Permalloy (Py) nanodot of $2R = 300$ nm diameter and $L = 10$ nm thickness [Fig. 1(a)]. For the given Py material and circular dot geometry, a single MV is present with either up- or down-core orientation and with either CCW or CW in-plane $\mathbf{M}$s around its core. The characteristic eigenfrequency [5] and the static vortex annihilation field [21] of the vortex were estimated to be $\nu_D = \omega_D/2\pi = 330$ MHz and $H_A = 500$ Oe, respectively [7]. We considered the application of either LPHs applied along the $y$ axis, $\mathbf{H}_{Lin} = H_0 \sin(\omega_\mathbf{H} t)\hat{\mathbf{y}}$ or circularly polarized oscillating fields (CPHs) of either CCW or CW rotation in the dot plane, such that $\mathbf{H}_{CCW,CW} = \pm H_0 \cos(\omega_\mathbf{H} t)\hat{\mathbf{x}} + H_0 \sin(\omega_\mathbf{H} t)\hat{\mathbf{y}}$, where $\omega_\mathbf{H}$ is the angular frequency and $H_0$ is the field amplitude. We used relatively low amplitudes, $H_0 / H_A = 0.1$ and 0.2, in investigations of only linear-regime vortex gyrotropic motions, so as to exclude the nonlinear effect [8] driven by high-strength fields. We also chose a frequency range from 100 to 825 MHz, including the vortex eigenfrequency of interest, $\nu_D = 330$ MHz, in order to investigate both the on- and off-resonance vortex gyrotropic motions.

First, we have to note that the orbital trajectory of the $\mathbf{H}_{Lin}$-driven steady-state VC motion for the resonance case $\omega_\mathbf{H} = \omega_D$ is exactly circular in shape and relatively large in amplitude, even for a very weak field strength (e.g. $H_0 / H_A = 0.01$ [22]), as shown in Figs. 1(b) and 1(c). For $p$

= +1, the VC motion is CCW and for $p$ = -1, CW. In the case of off-resonance ($\omega_H \neq \omega_D$) motion, the orbits, however, become elongated along the axis perpendicular to and parallel with the $\mathbf{H}_{Lin}$ axis for $\omega_H < \omega_D$ and $\omega_H > \omega_D$, respectively, and the degree of their elongations increases with increasing $|\omega_H - \omega_D|$ [8], as revealed by the resultant orbital trajectories shown in Fig. 1(d). Although such behaviors have been reported from numerical and analytical studies [7,8], the underlying physics has been incompletely understood. To our knowledge there have been no experimental data.

For clear understanding, it is thus necessary to find out the elementary eigenmodes of the gyrotropic motions. The application of the $\mathbf{H}_{Lin}$ is equivalent to the application of both the pure circular fields of $\mathbf{H}_{CCW}$ and $\mathbf{H}_{CW}$ with the same amplitude and frequency, as seen in Fig. 2(a), since an $\mathbf{H}_{Lin}$ with a $\omega_H$ can, in principle, be decomposed into $\mathbf{H}_{CCW}$ and $\mathbf{H}_{CW}$ rotating with the same $\omega_H$ and with equal $H_0$ [14]. The relative phase between $\mathbf{H}_{CCW}$ and $\mathbf{H}_{CW}$ determines the axis of an LPH. Thus, the observed circular or elliptical shape of the orbital trajectories of the linear-regime gyrotropic motions [Figs. 1(b), 1(d)] can be interpreted according to the superposition of the CCW and CW *circular* (elliptical) eigenmotions in the case of *circular* (elliptical) dots and with respect to the $\mathbf{H}_{CCW}$ and $\mathbf{H}_{CW}$ eigenbasis [Fig. 2(a)] [14]. To verify this by micromagnetic simulations, in Fig. 2(b) we plot the individual orbital trajectories of the VC motions under the individual components of $\mathbf{H}_{CCW}$ and $\mathbf{H}_{CW}$, as well as under the $\mathbf{H}_{Lin}$ for the case of $\omega_H / \omega_D = 2.5$, $H_0 / H_A = 0.2$, and $(p, C) = (+1, +1)$, where $C$ is the chirality, $C = +1$ ($C = -1$) for the CCW (CW) in-plane **M** rotation. The elliptical trajectory (black open circle) by the $\mathbf{H}_{Lin}$ is in excellent agreement with that (black solid circle) obtained by the superposition of the CCW (blue open circle) and CW (red open circle) circular-rotational motions.

In order to analytically verify/understand such dynamic responses of a vortex to any polarized oscillating field (or current), now we introduce a useful quantity of the dynamic susceptibility tensor $\hat{\chi}_\mathbf{X}$ defined by $\mathbf{X} = \hat{\chi}_\mathbf{X}\mathbf{H}$ (Ref. [10]). For convenience, first let us define the orbital radius $R_{CCW,CW}$ and phase $\delta_{CCW,CW}^\mathbf{H}$ of the VC position $\mathbf{X}$ in the dot ($x$-$y$) plane for the CCW and CW circular modes [Fig. 2(a)]. Here, we exclude non-steady transient-state motions that have yet to reach the steady state, as well as the nonlinear effect [8]. To analytically calculate $R_{CCW(CW)}$ and $\delta_{CCW(CW)}^\mathbf{H}$, we employed the linearized Thiele's equation [20] of motion, $-\mathbf{G}\times\dot{\mathbf{X}} - \hat{D}\dot{\mathbf{X}} + \partial W(\mathbf{X},t)/\partial \mathbf{X} = 0$, with the gyrovector $\mathbf{G} = -G\hat{\mathbf{z}}$, and the damping tensor $\hat{D} = D\hat{I}$ with the identity matrix $\hat{I}$ and the damping constant $D$ (Refs. [5],[10]). The potential energy function is given by $W(\mathbf{X},t) = W(0) + \kappa|\mathbf{X}|^2/2 + W_\mathbf{H}$. The first term $W(0)$ is the potential energy for a VC at its initial position $\mathbf{X} = 0$, and the second term is dominated by the exchange and magnetostatic energies for the VC shift from $\mathbf{X} = 0$ and for the given stiffness coefficient $\kappa$. The last one, $W_\mathbf{H} = -\mu(\hat{\mathbf{z}}\times\mathbf{H})\cdot\mathbf{X}$, is the Zeeman energy term due to a driving force, where $\mu = \pi R L M_s \xi C$ with $\xi = 2/3$ for the "side-charge-free" model [5,8,10]. For any polarized oscillating field (LPH, CPH, or their mixed polarizations), $\mathbf{H} = \mathbf{H}_0\exp(-i\omega_\mathbf{H}t)$, the general solution is $\mathbf{X} = \mathbf{X}^{trans} + \mathbf{X}^{steady}$, where $\mathbf{X}^{trans} = -\mathbf{X}_0\exp(-i\omega_D t)\exp(D\omega_D t/|G|)\mathbf{X}$ with $\omega_D = \kappa|G|/(G^2 + D^2)$ [10] and $\mathbf{X}^{steady} = \mathbf{X}_0 \exp(-i\omega_\mathbf{H} t)$. The terms $\mathbf{X}^{trans}$ and $\mathbf{X}^{steady}$ correspond to the VC motions in the transient- [8] and field-driven steady states, respectively.

For $t \gg |G/(D\omega_D)|$ ($t \gg 23$ ns in this case), the VC motions can be represented by $\mathbf{X} \simeq \mathbf{X}^{steady} = \mathbf{X}_0 \exp(-i\omega_\mathbf{H} t)$. For the LPH basis, the susceptibility tensor is $\mathbf{X}_{0,L} = \hat{\chi}_{\mathbf{X},L} \mathbf{H}_{0,L}$ with $\mathbf{X}_{0,L} = X_{0x}\hat{\mathbf{x}} + X_{0y}\hat{\mathbf{y}}$ and $\mathbf{H}_{0,L} = H_{0x}\hat{\mathbf{x}} + H_{0y}\hat{\mathbf{y}}$, where

$$\hat{\chi}_{X,L}(\omega_\mathbf{H}) = \begin{bmatrix} \chi_{xx} & \chi_{xy} \\ \chi_{yx} & \chi_{yy} \end{bmatrix} = \frac{\mu}{(i\omega_\mathbf{H} D + \kappa)^2 - (\omega_\mathbf{H} G)^2} \begin{bmatrix} -i\omega_\mathbf{H} G & -i\omega_\mathbf{H} D - \kappa \\ i\omega_\mathbf{H} D + \kappa & -i\omega_\mathbf{H} G \end{bmatrix}. \tag{1}$$

By the diagonalization of $\hat{\chi}_X$, $\mathbf{X}_{0,L} = \hat{\chi}_{\mathbf{X},L} \mathbf{H}_{0,L}$ becomes $\mathbf{X}_{0,cir} = \hat{\chi}_{\mathbf{X},cir} \mathbf{H}_{0,cir}$ with respect to the CPH eigenbasis, where $\mathbf{H}_{0,cir} = H_{0,CCW}\hat{\mathbf{e}}_{CCW} + H_{0,CW}\hat{\mathbf{e}}_{CW}$ and $\mathbf{X}_{0,cir} = X_{0,CCW}\hat{\mathbf{e}}_{CCW} + X_{0,CW}\hat{\mathbf{e}}_{CW}$ with the circular eigenvectors of $\hat{\mathbf{e}}_{CCW} = \frac{1}{\sqrt{2}}(\hat{\mathbf{x}} + i\hat{\mathbf{y}})$ and $\hat{\mathbf{e}}_{CW} = \frac{1}{\sqrt{2}}(\hat{\mathbf{x}} - i\hat{\mathbf{y}})$ [23]. $\mathbf{X}_{0,cir} = \hat{\chi}_{\mathbf{X},cir} \mathbf{H}_{0,cir}$ can also be rewritten, in matrix form, as $\begin{pmatrix} X_{0,CCW} \\ X_{0,CW} \end{pmatrix} = \begin{pmatrix} \chi_{CCW} & 0 \\ 0 & \chi_{CW} \end{pmatrix} \begin{pmatrix} H_{0,CCW} \\ H_{0,CW} \end{pmatrix}$, where $\chi_{CCW,CW} = i\mu/[\omega_\mathbf{H} G \mp (i\omega_\mathbf{H} D + \kappa)]$. The terms $\chi_{CCW,CW}$ can be separated into the magnitude $|\chi_{CCW,CW}|$ and phase $\delta^\mathbf{H}_{CCW,CW}$ by $\chi_{CCW,CW} = |\chi_{CCW,CW}| e^{-i\delta^\mathbf{H}_{CCW,CW}}$, where

$$|\chi_{CCW,CW}| = |\mu|/\sqrt{(G^2 + D^2)(\omega_\mathbf{H} \mp p\omega_D)^2 + \kappa^2 D^2/(G^2 + D^2)}, \tag{2a}$$

$$\delta^\mathbf{H}_{CCW,CW} = -\tan^{-1}[(\kappa \mp p\omega_\mathbf{H}|G|)/(\omega_\mathbf{H} D)] + \tfrac{\pi}{2}(1 \mp C). \tag{2b}$$

Both $|\chi_{CCW,CW}|$ and $\delta^\mathbf{H}_{CCW,CW}$ are functions of $\omega_\mathbf{H}$. $|\chi_{CCW,CW}|$ depends on the sign of $p$, independently of $C$ because of $|\mu| = \pi R L M_s \xi |C|$, whereas $\delta^\mathbf{H}_{CCW,CW}$ depends on the sign of both $p$ and $C$.

Next, the numerical solutions of the analytically derived $|\chi_{\text{CCW,CW}}|$ and $\delta^{\mathbf{H}}_{\text{CCW,CW}}$ were plotted in Fig. 3 versus $\omega_{\mathbf{H}}/\omega_D$ for four different cases of $(p, C)$, and were compared with the simulation results (circle symbols) according to $|\chi_{\text{CCW(CW)}}| = R_{\text{CCW(CW)}} / |\mathbf{H}_{\text{CCW(CW)}}|$, where $R_{\text{CCW(CW)}} \equiv |\mathbf{X}^{steady}_{\text{CCW(CW)}}|$. Both of the analytical results are in excellent agreement with the simulation results. There exist strong resonances for both $|\chi_{\text{CCW,CW}}|$ and $\delta^{\mathbf{H}}_{\text{CCW,CW}}$ at $\omega_{\mathbf{H}}/\omega_D = 1$ and the resonance effects are asymmetric between the CCW and CW circular motions. Only one, either the CCW or CW motion, shows a resonance behavior, the other showing non-resonance (compare sharp peak vs straight line), and the mode showing the resonance effect switches by the $p$ of the given vortex. This asymmetric resonance effect can be ascribed to the presence of a VC of either $p = +1$ or $p = -1$, and is clearly, analytically understood with reference to the on-resonance case equations, $|\chi_{\text{CCW}}| = (|\mu|/\kappa)\sqrt{(G^2 + D^2)/D^2}$ and $|\chi_{\text{CW}}| = (|\mu|/\kappa)\sqrt{(G^2 + D^2)/(4G^2 + D^2)}$, which yield $|\chi_{\text{CCW}}| \gg |\chi_{\text{CW}}|$ for $p = +1$. Alternatively, for $p = -1$, $|\chi_{\text{CCW}}| = (|\mu|/\kappa)\sqrt{(G^2 + D^2)/(4G^2 + D^2)}$ and $|\chi_{\text{CW}}| = (|\mu|/\kappa)\sqrt{(G^2 + D^2)/D^2}$, thus yielding $|\chi_{\text{CCW}}| \ll |\chi_{\text{CW}}|$. These analytical interpretations are also well verified by their numerical calculations (solid lines) [Fig. 3]. The asymmetric resonance effect is caused by the gyroforce ($\mathbf{G} \times \dot{\mathbf{X}}$), which is essential for vortex gyrotropic motion. The presence of the gyroforce leads to a broken time-reversal symmetry, in turn yielding a splitting of the degeneracy of the CCW and CW eigenmodes. Therefore, the vortex gyrotropic motion shows such asymmetric resonance, responding differently to the orthogonal CCW and CW circular fields, and the asymmetric resonance effect is reversed by changing from $p = +1$ to $p = -1$.

Compared with the variations of $|\chi_{\text{CCW,CW}}|$ with $\omega_\text{H}$, the $\delta^\text{H}_{\text{CCW,CW}}$ variation with $\omega_\text{H}$ is more remarkable, owing to the $C$ as well as $p$ dependences of the $\delta^\text{H}_{\text{CCW,CW}}$. The $C$ dependence originates from applied driving forces, $\mathbf{F}^\text{H}_{\text{CCW,CW}} = \mu(\hat{\mathbf{z}} \times \mathbf{H}_{\text{CCW,CW}}) = |\mu|e^{\mp iC\pi/2}\mathbf{H}_{\text{CCW,CW}}$, because the sign of $\mu$ changes with $C$. The $p$ dependence, meanwhile, is due to the already-mentioned asymmetric resonance effect between $|\chi_{\text{CCW}}|$ and $|\chi_{\text{CW}}|$. As in the dynamic response of a linear oscillator to a harmonic oscillating force, for $\omega_\text{H} < \omega_D$ the phase difference $\delta^\text{F}_{\text{CCW,CW}}$ between the VC position $\mathbf{X}$ and $\mathbf{F}^\text{H}_{\text{CCW,CW}}$ is always zero (i.e., in phase), and independent of $C$ and $p$. However, for the other case, $\omega_\text{H} > \omega_D$, $\delta^\text{F}_{\text{CCW,CW}} = -\pi(1 \pm p)/2$ depends only on $p$. Only for the case of the eigenmode showing resonance, $\delta^\text{F}_{\text{CCW,CW}}$ changes from 0 (in-phase) at $\omega_\text{H} < \omega_D$ to $-\pi$ (out-of-phase) at $\omega_\text{H} > \omega_D$, as shown in the second row of Fig. 3. In addition, it is worthwhile noting that such phase change with $\omega_\text{H}$ takes place only for the eigenmode showing resonance; it does not occur for the other opposite eigenmode. Thus, the complex changes of $\delta^\text{H}_{\text{CCW,CW}}$ with $\omega_\text{H}$, depending on $p$ and $C$, can be interpreted simply according to $\delta^\text{H}_{\text{CCW,CW}} = \delta^\text{F}_{\text{CCW,CW}} \pm C\pi/2$, as seen in the third row of Fig. 3.

According to the above-mentioned results, the CCW and CW circular motions are expressed individually by $\mathbf{X}_{\text{CCW}} = \chi_{\text{CCW}}\mathbf{H}_{\text{CCW}}$ and $\mathbf{X}_{\text{CW}} = \chi_{\text{CW}}\mathbf{H}_{\text{CW}}$, respectively, where $\mathbf{H}_{\text{CCW}} = H_{0,\text{CCW}}\exp(-i\omega_\text{H}t)\hat{\mathbf{e}}_{\text{CCW}}$ and $\mathbf{H}_{\text{CW}} = H_{0,\text{CW}}\exp(-i\omega_\text{H}t)\hat{\mathbf{e}}_{\text{CW}}$. Consequently, the superposition of the individual circular eigenmotions thus provides a resultant VC gyrotropic motion driven by an $\mathbf{H}_{\text{Lin}}(\omega_{\mathbf{H}})$ [24]. For concrete verification, the individual orbital trajectories of the CCW and CW eigenmodes in response to the $\mathbf{H}_{\text{CCW}}$ and $\mathbf{H}_{\text{CW}}$ eigenbasis and their superposition are also obtained from the analytical calculations (solid lines), for example for the different cases of $\omega_{\mathbf{H}}/\omega_D = 0.3 < 1$ and $2.5 > 1$ with $H_0/H_A = 0.2$ and for $(p, C) = (+1, +1)$, as seen in Fig. 4(a). The other cases of $(p, C)$ are shown in Supplementary Fig. 2 [25]. The analytical calculations are in excellent agreement with the simulation results (open circles). The elongation of the orbits for $\omega_{\mathbf{H}}/\omega_D = 0.3$ and 2.5 shows almost the same. By contrast, $\delta_{\text{CCW}}^{\mathbf{H}} \approx \pi/2$ and $-\pi/2$ for $\omega_{\mathbf{H}}/\omega_D = 0.3$ and 2.5, respectively, but $\delta_{\text{CW}}^{\mathbf{H}} \approx -\pi/2$ for both cases. It is sure that the degree of elongation and rotation of the orbital trajectories of those vortex motions shown in Fig. 1(d) are related closely to the $|\chi_{\text{CCW,CW}}|$ and $\delta_{\text{CCW,CW}}^{\mathbf{H}}$ responses to the different values of $\omega_{\mathbf{H}}$, depending on $p$ and $C$. For more quantitative understanding, it is convenient to define the ellipticity $\eta_{\mathbf{G}}$ as the ratio of the length of the major (*a*) to that of the minor (*b*) axis, and the rotation $\theta_{\mathbf{G}}$ as the angle of the ellipse's major axis from the LPH axis (the *y* axis in this case) [Fig. 4(b)], as in the Kerr or Faraday ellipticity and rotation in magneto-optics [24]. The numerical values of $\eta_{\mathbf{G}}$ and $\theta_{\mathbf{G}}$, which are $\eta_{\mathbf{G}} = (R_{\text{CCW}} - R_{\text{CW}})/(R_{\text{CCW}} + R_{\text{CW}})$ and $\theta_{\mathbf{G}} = (\delta_{\text{CCW}}^{\mathbf{H}} - \delta_{\text{CW}}^{\mathbf{H}})/2$ by definition, were plotted versus $\omega_{\mathbf{H}}/\omega_D$ in Fig. 4(c), for a case of $(p, C) = (+1, +1)$ [The open circle and square symbols indicate simulation data corresponding to the cases shown in the left and right columns of Fig. 1(d), respectively]. The $\eta_{\mathbf{G}}$ and $\theta_{\mathbf{G}}$ values for the other cases of (*p*, *C*) are shown in Supplementary Fig. 3 [25]. Note that the analytical calculations (solid lines) of $\eta_G$ and $\theta_G$ are in excellent agreement with the simulation results (symbols). Owing to the resonance characteristics of either the CCW or CW eigenmode for a given *p*, $\eta_G$ and $\theta_G$ dramatically change across $\omega_H/\omega_D = 1$ such that $\eta_G = +1$ or -1 (indicating that the orbital shape is circular) and $\theta_G = +\pi/4$ or $-\pi/4$ (Figs. 4(c) and Supplementary Fig. 3) [26]. The sign of $\eta_G$ represents the rotation sense of the resultant VC gyrotropic motion driven by an LPH (i.e., CCW and CW rotation for $\eta_G > 0$ and $\eta_G < 0$, respectively). Since the relative magnitude of $R_{CCW}$ and $R_{CW}$ is determined by *p*, not by *C*, (i.e. $R_{CCW} > R_{CW}$ for $p = +1$ and $R_{CCW} < R_{CW}$ for $p = -1$), the sign of $\eta_G$ is determined by *p*, e.g., $\eta_G < 0$ for $p = -1$ and $\eta_G > 0$ for $p = +1$. By contrast, $\theta_G$ depends on both *p* and *C* because of their dependences of $\delta_{CCW,CW}^H$. The sharp variation of $\theta_G$ from $\pi/2$ to 0 with increasing $\omega_H$ across $\omega_D$ indicates that the major axis of the ellipses changes from the *x* axis to the *y* axis. From the calculations of $\delta_{CCW,CW}$ for each *p* and *C*, we can expect that $\theta_G = +\pi/2$ or $-\pi/2$ (the major axis is perpendicular to the $\mathbf{H}_{Lin}$ axis) for $\omega_H < \omega_D$, and that $\theta_G = 0$ (the major axis is parallel to the $\mathbf{H}_{Lin}$ axis) for $\omega_H > \omega_D$, regardless of *p* and *C* [see Fig. 4(c) and Supplementary Fig. 3].

In conclusion, we found that the CCW and CW circular-rotational eigenmodes in the vortex gyrotropic motions with respect to the corresponding circular-rotational fields and their asymmetric resonant excitations are instrumental to the understanding of the dependences of the elongation and orientation of the orbital trajectories of the linear-regime steady-state motions on not only the frequency of any linearly polarized in-plan oscillating fields but also the polarization and chirality of the given vortex.

*Note added*: We became aware of the presentation HE-02 at the 52nd annual conference on Magnetism and Magnetic Materials in Tampa, Florida in 2007, as well as the manuscript by B. Kruger *et al*. [*Phys. Rev. B* **76** 224426 (2007)].

## Acknowledgements

We thank Dae-Eun Jeong, Young-Sang Yu, Yun-Seok Choi, and K. Yu Guslienko for their valuable assistances. This work was supported by Creative Research Initiatives (Research Center for Spin Dynamics and Spin-Wave Devices) of MOST/KOSEF.

# References

*Corresponding author: sangkoog@snu.ac.kr

modes of left- and right-handed circular polarizations, which have different amplitudes and phases owing to their different refractive indices in a given ferromagnetic medium.

[25] See EPAPS Document No. XXX for three supplementary figures and two movie files. For more information on EPAPS, see http://www.aip.org/pubservs/epaps.html.

[26] Here, we emphasize only the linear regime for relatively low values of $H_0 / H_A$. As far as the linear regime is considered, the values of $\eta_G$ and $\theta_G$ do not change with $H_0 / H_A$.

**Figure captions**

FIG. 1. (color online) (a) Geometry and dimension of the model Py nanodot with a vortex-state **M** distribution with the up-core orientation and CCW in-plane **M** rotation around its VC at equilibrium under no field. (b) Perspective view of the local **M** distributions at the indicated times and (c) the circular orbital trajectory of the VC motion in the steady state, driven by $\mathbf{H}_{\text{Lin}} = H_0 \sin(\omega_\mathbf{H} t)\hat{\mathbf{y}}$ with $H_0/H_A = 0.01$ and $\omega_\mathbf{H}/\omega_D = 1$. The color and height display the local in-plane **M** orientation, as indicated by the color wheel, as well as the out-of-plane **M** components, respectively. The dots in (c) indicate the VC positions at the indicated times. (d) Orbital trajectories of the steady-state VC motions ($t > 90$ ns) in response to the $\mathbf{H}_{\text{Lin}}$ with different $\omega_\mathbf{H}$ values as noted for $H_0/H_A = 0.1$ (left column) and 0.2 (right column).

FIG. 2. (color online) (a) Graphical illustration of the CCW and CW circular eigenmodes and the corresponding circular eigenbasis of $\mathbf{H}_{\text{CCW}}$ and $\mathbf{H}_{\text{CW}}$, as well as the definitions of the amplitude $R_{\text{CCW,CW}}$ and phase $\delta^{\mathbf{H}}_{\text{CCW,CW}}$ of the linear-regime steady-state circular VC motions. The black-colored ellipse indicates the resultant superposition of the individual CCW and CW eigenmodes, which is equivalent to the elliptical VC motion driven by the $\mathbf{H}_{\text{Lin}}$ (black arrow), the sum of $\mathbf{H}_{\text{CCW}}$ and $\mathbf{H}_{\text{CW}}$. The VC positions of the CCW and CW eigenmotions at a certain time are indicated by the blue- and red-colored circles, respectively, and their vector sum is indicated by the black-colored circle on the ellipse. (b) Micromagnetic simulation results on the VC trajectories of the CCW (blue open circles) and CW (red open circles) eigenmotions driven by the individual $\mathbf{H}_{\text{CCW}}$ and $\mathbf{H}_{\text{CW}}$, respectively, as well as the VC trajectory (black open circles) of the motion driven by the $\mathbf{H}_{\text{Lin}}$ for ($\omega_{\mathbf{H}}/\omega_D$, $H_0/H_A$)=(2.5, 0.2) and ($p$, $C$) = (+1, +1). The elliptical orbit (black solid circles) results from the superposition of the individual CCW and CW eigenmotions.

FIG. 3. (color online) Numerical calculations of the analytical equations (solid lines) of $|\chi_{\text{CCW,CW}}| = R_{\text{CCW,CW}}/|\mathbf{H}_{\text{CCW,CW}}|$, $\delta^{\mathbf{F}}_{\text{CCW,CW}}$, and $\delta^{\mathbf{H}}_{\text{CCW,CW}}$, compared with the micromagnetic simulations (shaded circles) for the CCW and CW eigenmodes in response to the pure $\mathbf{H}_{\text{CCW}}$ (blue) and $\mathbf{H}_{\text{CW}}$ (red), respectively, versus $\omega_{\mathbf{H}}/\omega_D$ for the given polarization and chirality ($p$, $C$), as noted.

FIG. 4. (color online) (a) Simulation (open circles) and analytical (sold lines) calculations of the orbital trajectories of the VC motions driven by $\mathbf{H}_{\text{CCW}}$ (blue) and $\mathbf{H}_{\text{CW}}$ (red), as well as $\mathbf{H}_{\text{Lin}}$ (black) for $H_0/H_A = 0.2$ and ($p$, $C$) = (+1, +1) in the two cases of $\omega_{\mathbf{H}}/\omega_D = 0.3$ and 2.5. The phase relation between the VC position (closed dot) and the circular field direction (arrow) is illustrated. The arrows on the circular or elliptical orbits indicate their rotation senses. (b) Illustration of the definitions of $\theta_{\mathbf{G}}$ and $\eta_{\mathbf{G}}$ described in the text. (c) Numerical estimates of $\theta_{\mathbf{G}}$ and $\eta_{\mathbf{G}}$, obtained from the micromagnetic simulations (symbols) and numerical calculations of the analytical equations (solid lines), as a function of $\omega_H/\omega_D$ for a case of $(p, C) = (+1,+1)$. The circle and square symbols in (c) correspond to the results for the cases shown in the first and second columns in Fig. 1(d), respectively.

**FIG. 1.**

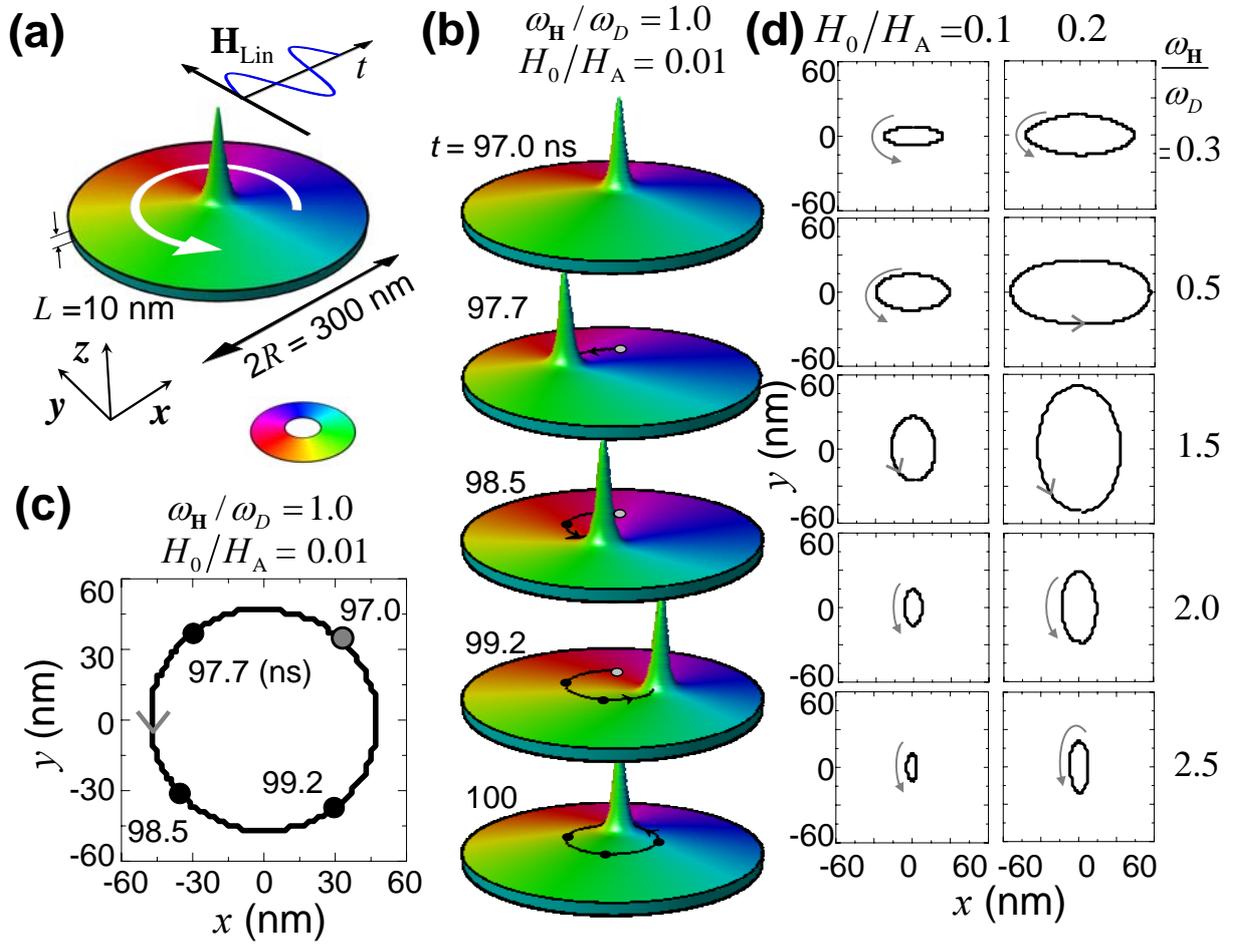

FIG. 2.

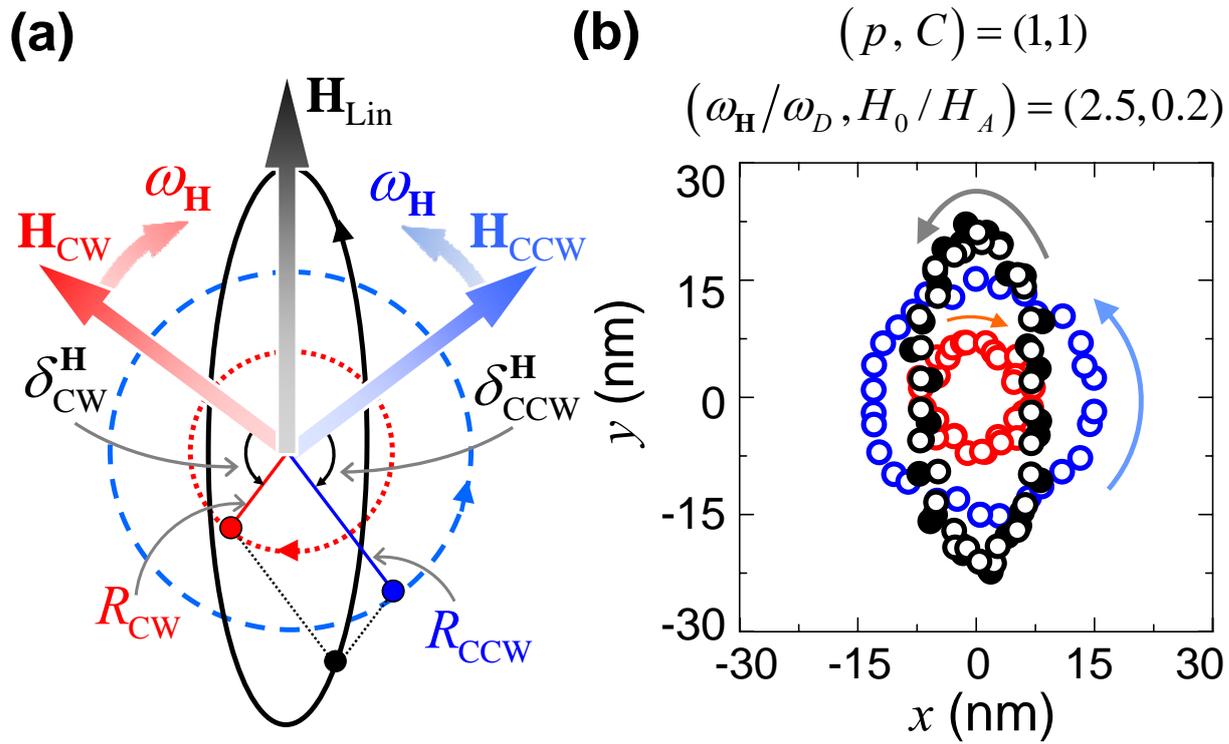

**FIG. 3.**

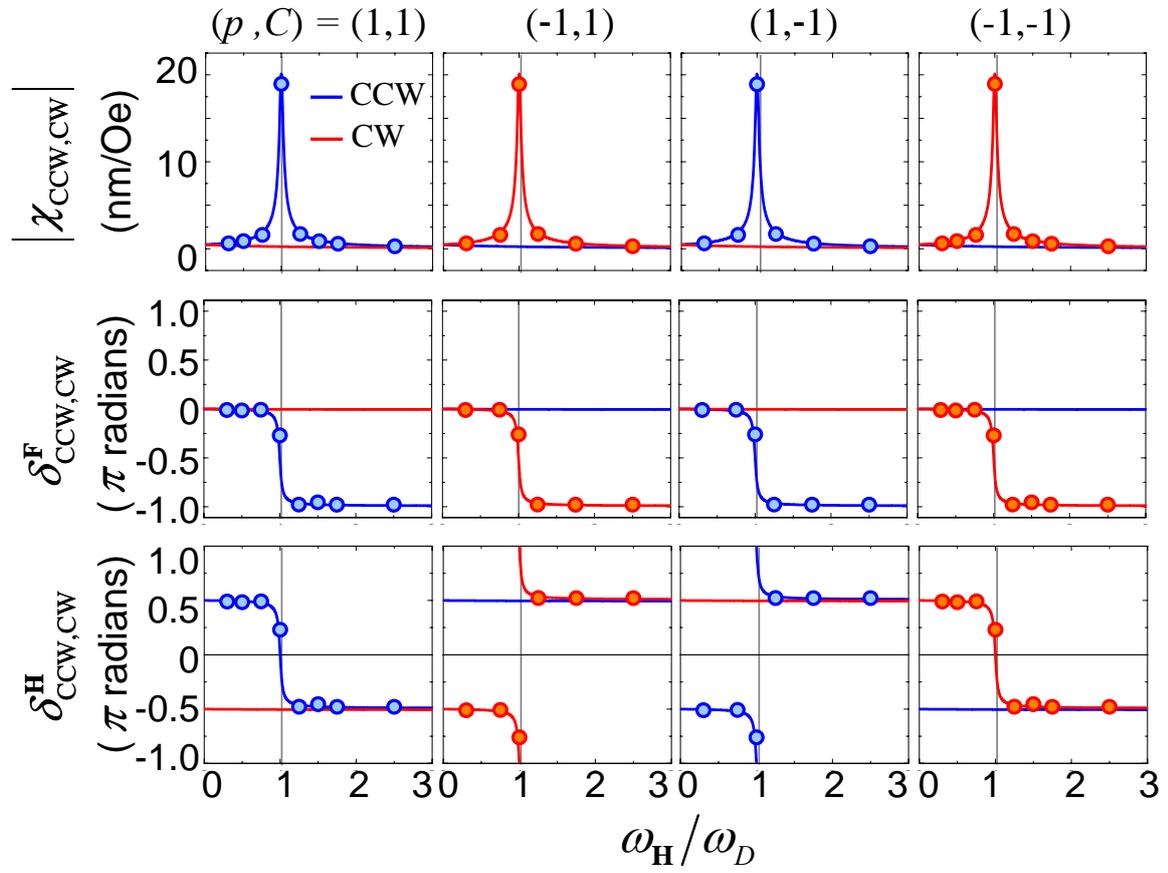

**FIG. 4.**

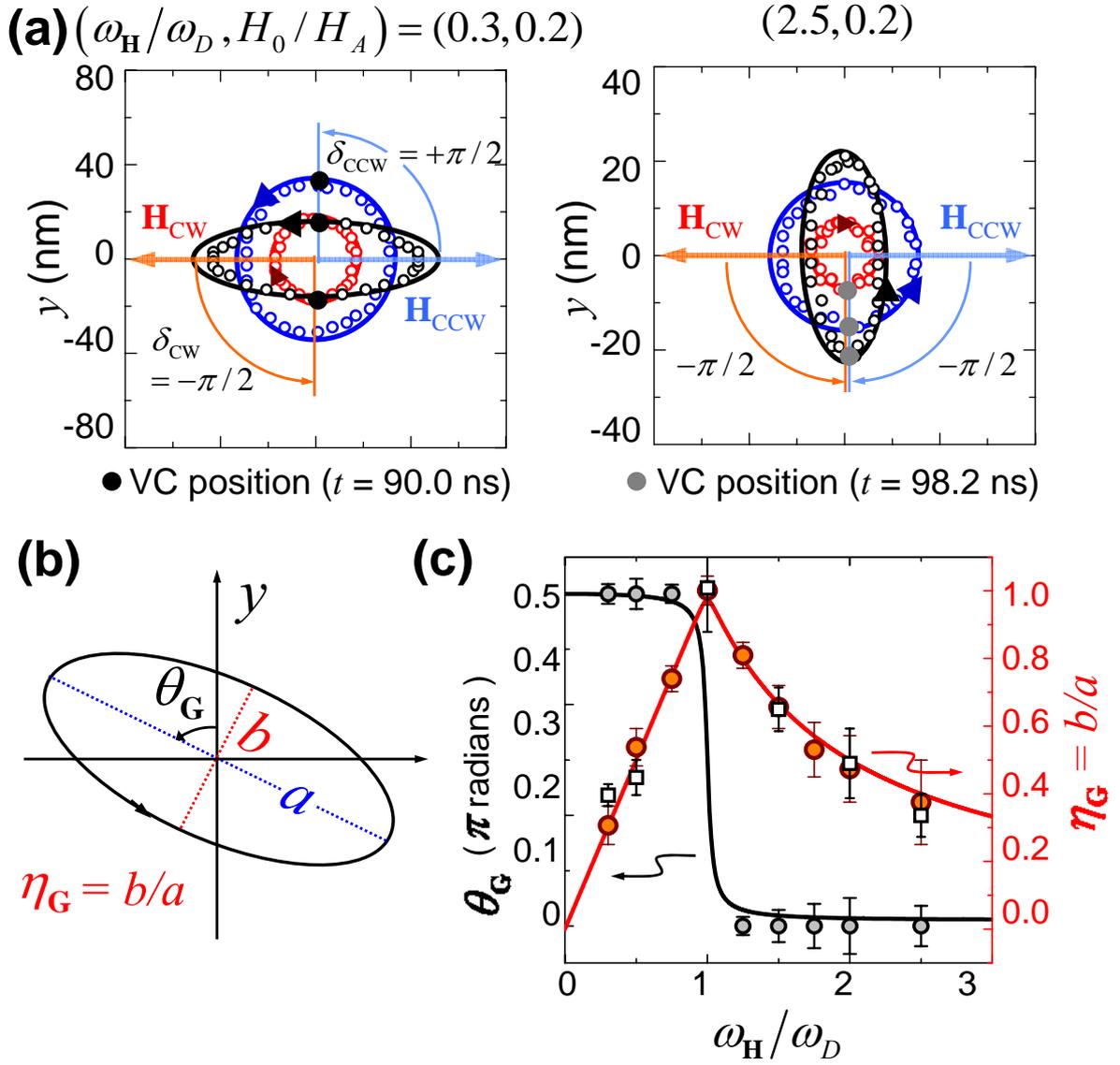

# Supplementary Movies

**Supplementary Movie 1.** Animation on the temporal evolution of the spatial configuration of the local magnetizations (**M**s) and of the VC motion for the case shown in Fig 1(c). Both the colors and the height of the surface indicate the local out-of-plane **M** component normalized by the saturation value.

**Supplementary Movie 2.** Animation on the temporal evolution of the individual CCW and CW circular eigenmodes and of the corresponding circular eigenbasis of the pure circular fields $\mathbf{H}_{CCW}$ and $\mathbf{H}_{CW}$, and with their superposition. The blue (red) line indicates the orbital trajectories of the steady-state CCW (CW) VC motion in the linear regime. The black ellipse indicates the resultant superposition of the individual CCW and CW eigenmodes, which is equivalent to the elliptical VC motion driven by the $\mathbf{H}_{Lin}$ (black arrow), the sum of $\mathbf{H}_{CCW}$ and $\mathbf{H}_{CW}$.

# Supplementary Figures and Captions

**SUPPLEMENTARY FIG. 1.**

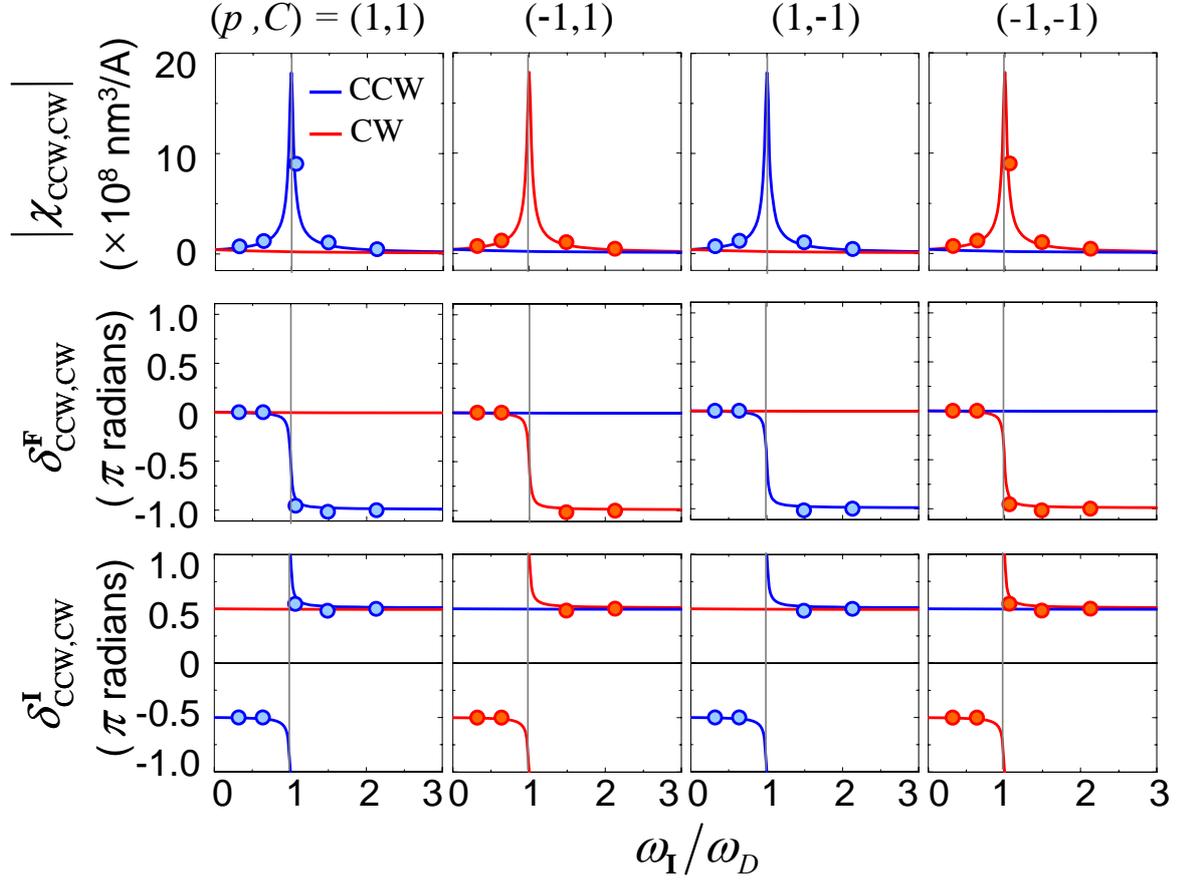

**Supplementary Figure 1.** Numerical calculations of the analytical equations (solid lines) of $\left|\chi^{\mathbf{I}}_{\text{ccw,cw}}\right| = R_{\text{ccw,cw}} / \left|\mathbf{j}_{\text{ccw,cw}}\right|$, $\delta^{\mathbf{I}}_{\text{ccw,cw}}$, and $\delta^{\mathbf{F}}_{\text{ccw,cw}}$, compared with the simulation results (symbols) for the CCW and CW eigenmodes in response to the pure $\mathbf{j}_{\text{ccw}}$ (blue) and $\mathbf{j}_{\text{cw}}$ (red) versus $\omega_{\mathbf{I}}/\omega_{D}$ for the given polarization and chirality ($p$, $C$), as noted.

**SUPPLEMENTARY FIG. 2.**

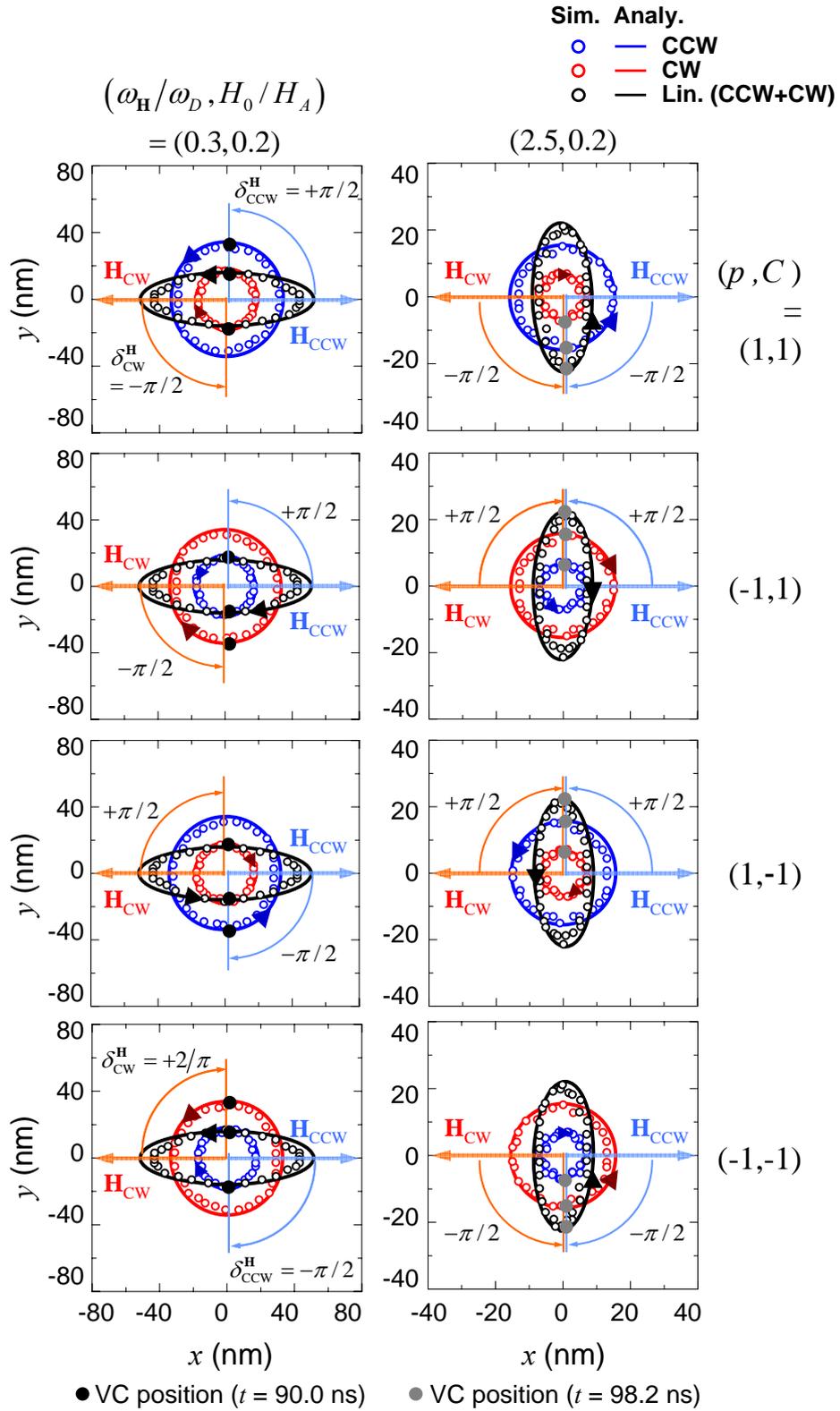

● VC position (*t* = 90.0 ns)   ● VC position (*t* = 98.2 ns)

**Supplementary FIG. 2.** Simulation (open circles) and analytical (solid lines) calculations of the orbital trajectories of the VC motions driven by $\mathbf{H}_{CCW}$ (blue) and $\mathbf{H}_{CW}$ (red), as well as $\mathbf{H}_{Lin}$ (black) for different cases of ($p$, $C$) as noted for $H_0 / H_A = 0.2$, and $\omega_H / \omega_D = 0.3$ and 2.5. The phase relation between the VC position (closed dot) and the circular field direction (arrow) is illustrated. The arrows on the circular or elliptical orbits indicate their rotation senses.

**SUPPLEMENTARY FIG. 3.**

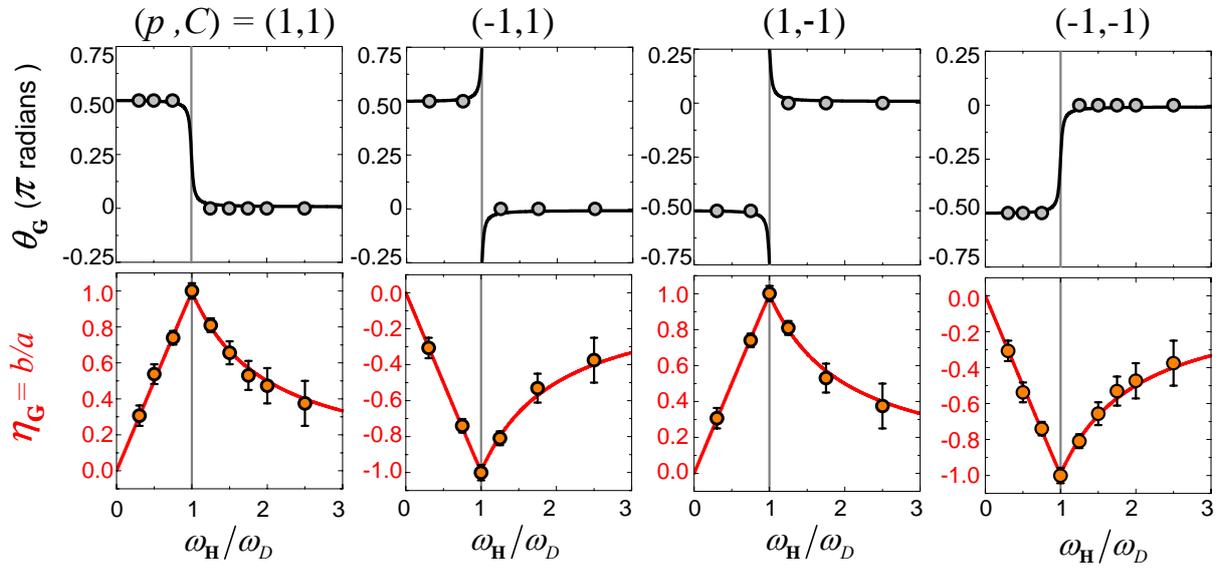

**Supplementary FIG.3.** Results of the micromagnetic simulations (symbols) and numerical calculations of the analytical equations (solid lines) of $\theta_G$ and $\eta_G$, as a function of $\omega_H/\omega_D$ for different cases of ($p$, $C$), as noted.

## More detail of supplementary Figure 1

It is also interesting to consider vortex gyrotropic motions driven by spin-polarized alternating currents. To derive the elementary CCW and CW circular eigenmodes in circular dots, it is convenient to consider the force, driven by spin-polarized currents, that is given by $\mathbf{F}^{\mathbf{I}} = -\mathbf{G} \times \mathbf{v}_s = p|G|\hat{\mathbf{z}} \times \mathbf{v}_s$, with the drift velocity of electron spins $\mathbf{v}_s = Pa^3 \mathbf{j}_s/(2eS)$, where $P$ is the spin polarization, $a$ is the lattice constant, $e$ is the absolute value of the electronic charge, $S$ is the magnitude of spin, and where the spin current density $\mathbf{j}_s = -\mathbf{j}$. The linearized Thiele's equation of the current driven vortex motion, including the spin-polarized currents term, is given as $\mathbf{G} \times (\dot{\mathbf{X}} - \mathbf{v}_s) + \hat{D}\dot{\mathbf{X}} - \kappa|\mathbf{X}| = 0$ [1]. By inserting the driving force of $-\mathbf{G} \times \mathbf{v}_s$, the equation of motion is finally given by $\mathbf{G} \times \dot{\mathbf{X}} + \hat{D}\dot{\mathbf{X}} - \kappa|\mathbf{X}| + \mu_I(\hat{\mathbf{z}} \times \mathbf{j}) = 0$, where $\mu_I = -p|G|Pa^3/(2eS)$. This equation of motion is similar to that driven by an oscillating magnetic field, where $\mu$ and $\mathbf{H}$ are replaced by $\mu_I$ and $\mathbf{j}$, respectively, in the case of currents. Consequently, the CCW and CW circular eigenmodes driven by the circular-rotational current basis are expressed as $\mathbf{X}_{CCW} = \chi^{\mathbf{I}}_{CCW}\mathbf{j}_{CCW}$ and $\mathbf{X}_{CW} = \chi^{\mathbf{I}}_{CW}\mathbf{j}_{CW}$ with $\mathbf{j}_{CCW} = j_{0,CCW}\exp(-i\omega_I t)\hat{\mathbf{e}}_{CCW}$ and $\mathbf{j}_{CW} = j_{0,CW}\exp(-i\omega_I t)\hat{\mathbf{e}}_{CW}$, where $\chi^{\mathbf{I}}_{CCW} = |\chi^{\mathbf{I}}_{CCW}|e^{-i\delta^{\mathbf{I}}_{CCW}} = i\mu_I/[\omega G - (i\omega D + \kappa)]$ and $\chi^{\mathbf{I}}_{CW} = |\chi^{\mathbf{I}}_{CW}|e^{-i\delta^{\mathbf{I}}_{CW}} = i\mu_I/[\omega G + (i\omega D + \kappa)]$.

The numerical calculations of $|\chi^{\mathbf{I}}_{CCW,CW}|$, $\delta^{\mathbf{I}}_{CCW,CW}$, and $\delta^{\mathbf{F}}_{CCW,CW}$ versus $\omega_H/\omega_D$ and their simulation results for different cases of ($p$, $C$) are in excellent agreement with the simulation results, as seen in Supplementary Fig. 1. There is the same asymmetric resonance effect between the CCW and CW modes as that for oscillating magnetic fields. In contrast to $\delta_{\mathrm{CCW,CW}}^{\mathrm{H}}$, which varies with both $C$ and $p$, $\delta_{\mathrm{CCW,CW}}^{\mathrm{I}}$ depends only on $p$. This is because the current-induced driving force exerting on the vortex does not depend on $C$, but rather on $p$, as represented by $\mathbf{F}_{\mathrm{CCW}}^{\mathrm{I}} = \mu_I (\hat{\mathbf{z}} \times \mathbf{j}_{\mathrm{CCW}}) = |\mu_I| \exp(+ip\pi/2) \mathbf{j}_{\mathrm{CCW}}$ and $\mathbf{F}_{\mathrm{CW}}^{\mathrm{I}} = \mu_I (\hat{\mathbf{z}} \times \mathbf{j}_{\mathrm{CW}}) = |\mu_I| \exp(-ip\pi/2) \mathbf{j}_{\mathrm{CW}}$, whereas $\mathbf{F}_{\mathrm{CCW,CW}}^{\mathrm{H}}$ depends on both $C$ and $p$. As a result, both values of $\theta_{\mathrm{G}}$ and $\eta_{\mathrm{G}}$, in the case of applied currents, depend only on $p$, unlike the case of applied fields.

**Reference**

[1] J. Shibata, Y. Nakatani, G. Tatara, H. Kohno, and Y. Otani, Phys. Rev. B **73**, 020403(R) (2006)

# More details of Supplementary Figure 2 and Figure 3

The individual orbital trajectories of the CCW and CW eigenmodes in response to the $\mathbf{H}_{\text{CCW}}$ and $\mathbf{H}_{\text{CW}}$ eigenbasis and their superposition are also obtained from the analytical calculations (solid lines), for example for the different cases of $\omega_{\mathbf{H}}/\omega_D = 0.3 < 1$ and $2.5 > 1$ with $H_0/H_A = 0.2$ and for all the cases of ($p$, $C$), as seen in Supplementary Fig. 2. The analytical calculations (solid lines) are in excellent agreement with the simulation results (open circles). The relations of $\delta^{\mathbf{H}}$ and $R$ between CCW and CW circular motions determine $\eta_G$ and $\theta_G$, respectively. For each case of ($p$, $C$), those results and shown in Suppl. Table 1 for two different cases of $\omega_{\mathbf{H}}/\omega_D$ =0.3 and 2.5.

**Supplementary Table 1**

| ($p$, $C$) | (1, 1) | | (-1, 1) | | (1, -1) | | (-1, -1) | |
|---|---|---|---|---|---|---|---|---|
| $\omega_{\mathbf{H}}/\omega_D$ | 0.3 | 2.5 | 0.3 | 2.5 | 0.3 | 2.5 | 0.3 | 2.5 |
| $\delta_{\text{CCW}}^{\mathbf{H}}$ | $\frac{\pi}{2}$ | $-\frac{\pi}{2}$ | $\frac{\pi}{2}$ | $\frac{\pi}{2}$ | $-\frac{\pi}{2}$ | $\frac{\pi}{2}$ | $-\frac{\pi}{2}$ | $-\frac{\pi}{2}$ |
| $\delta_{\text{CW}}^{\mathbf{H}}$ | $-\frac{\pi}{2}$ | $-\frac{\pi}{2}$ | $-\frac{\pi}{2}$ | $\frac{\pi}{2}$ | $\frac{\pi}{2}$ | $\frac{\pi}{2}$ | $\frac{\pi}{2}$ | $-\frac{\pi}{2}$ |
| $R_{\text{CCW}}/R_{\text{CW}}$ | 1.86 | 2.33 | 0.54 | 0.43 | 1.86 | 2.33 | 0.54 | 0.43 |
| $\eta_G$ | 0.3 | 0.4 | -0.3 | -0.4 | 0.3 | 0.4 | -0.3 | -0.4 |
| $\theta_G$ | $\frac{\pi}{2}$ | 0 | $\frac{\pi}{2}$ | 0 | $-\frac{\pi}{2}$ | 0 | $-\frac{\pi}{2}$ | 0 |